# Dynamic realization of emergent high-dimensional optical vortices

Dongha Kim[1,†,#], Geonhyeong Park[2,#], Yun-Seok Choi[3], Arthur Baucour[4], Jisung Hwang[4], Sanghyeok Park[5], Hee Seong Yun[2], Jonghwa Shin[4], Haiwen Wang[6], Shanhui Fan[1], Dong Ki Yoon[2,‡], and Min-Kyo Seo[5,§]

[1]Ginzton Laboratory, Stanford University, Via Pueblo 348, CA, 94305, USA

[2]Department of Chemistry, KAIST, Daehak-ro, 291, 34141, Republic of Korea

[3]Center for Integrated Nanotechnologies, Los Alamos National Laboratory, Los Alamos, NM, 87545 USA

[4]Department of Materials Science and Engineering, KAIST, Daehak-ro, 291, 34141, Republic of Korea

[5]Department of Physics, KAIST, Daehak-ro, 291, 34141, Republic of Korea

[6]Department of Applied Physics, Stanford University, Via Pueblo 348, CA, 94305, USA

#These authors equally contributed to this work.

†dongha@stanford.edu, ‡nandk@kaist.ac.kr, §minkyo_seo@kaist.ac.kr

## Abstract

The dimensionality of vortical structures has recently been extended beyond two dimensions, providing higher-order topological characteristics and robustness for high-capacity information processing and turbulence control. The generation of high-dimensional vortical structures has mostly been demonstrated in classical systems through the complex interference of fluidic, acoustic, or electromagnetic waves. However, natural materials rarely support three- or higher-dimensional vortical structures and their physical interactions. Here, we present a high-dimensional gradient thickness optical cavity (GTOC) in which the optical coupling of planar metal-dielectric multilayers implements topological interactions across multiple dimensions. Topological interactions in high-dimensional GTOC construct non-trivial topological phases, which induce high-dimensional vortical structures in generalized parameter space in three, four dimensions, and beyond. These emergent high-dimensional vortical structures are observed under electro-optic tomography as optical vortex dynamics in two-dimensional real-space, employing the optical thicknesses of the dielectric layers as synthetic dimensions. We experimentally demonstrate emergent vortical structures, optical vortex lines and vortex rings, in a three-dimensional generalized parameter space and their topological transitions. Furthermore, we explore four-dimensional vortical structures, termed optical vortex sheets, which provide the programmability of real-space optical vortex dynamics. Our findings hold significant promise for emulating high-dimensional physics and developing active topological photonic devices.

**Introduction**

As topological textures, vortices possess the orbital nature of physical entities and have been extensively explored in a broad range of scientific disciplines [1]. The generation and manipulation of vortices have been demonstrated in both classical and quantum systems, ranging from fluidic [2], acoustic [3], photonic [4,5], magnetic [6], and electronic systems [7,8]. Most of classical systems consist of non-interactive media, which require a rotational perturbation for vortex generation [3,4,9]. In interactive or quantum systems, exotic phases of matter spontaneously generate vortices as quasiparticles in natural materials [7,8,10,11,12]. The interaction between spontaneously generated vortices enables us to explore not only novel quantum interactions, such as quantum turbulence decay [13,14] and non-Abelian anyons [15], but also the unique phases of matter formed through vortex-vortex interactions, such as vortex matter [16,17] and the Berezinskii-Kosterlitz-Thouless transition [7,18]. Like the optical analogues of other quantum phenomena and relativistic effects, including the photonic spin Hall effect [19,20] and the optical black hole [21,22], the classical emulation of the rich physics provided by spontaneously generated vortices will be an important inspiration for related research. Early photonic platforms for exploring the underlying physics of vortex phenomena, such as a photorefractive crystal [23] and a gradient-thickness optical cavity (GTOC) [24], have recently been developed.

Meanwhile, the exploration of vortices has been recently extended from two-dimensional (2-D) to three-dimensional (3-D) spaces. By spatial engineering of phase retardation elements, 3-D vortical structures, such as vortex lines, rings, and knots, have been generated by complex interferences of classical waves, which have been conducted in fluidic [25], acoustic [26], and photonic systems [27]. Vortex rings and knots cannot be effectively described in a low-dimensional space due to their higher-order topology compared to simple 2-D vortices. The robustness and scalability of the topological characteristics enable the application of 3-D vortical structures towards high-dimensional information storage and processing [28,29,30]. Furthermore, the collisions between the vortical structures can create complex swirling textures in 3-D space, providing an extra degree of freedom in the control of turbulent phenomena [31,32,33].

However, generation of 3-D vortical structures has been rarely reported in natural materials due to the lack of high-dimensional topological interaction. Examples of 3-D vortical structures in condensed matter platforms include vortex rings in liquid crystal, magnetic, and atomic media [34,35,36,37] and vortex line lattices in superconducting media [16,17], but their generations are probabilistic. Therefore, the implementation of high-dimensional topological interaction in physical systems will lead to the deterministic generation of higher-order topological textures, which provides multiple degrees of freedom for the manipulation of various physical entities. Recently, novel photonic platforms have enabled the investigation of high-dimensional physics in lower-dimensional systems through the use of synthetic dimensions [38,39,40,41,42,43,44]. In this study, we explore emergent optical vortices (OVs) with high-order topology in the GTOC platform by introducing synthetic spatial dimensions.

**High-dimensional GTOC**

We propose a topological platform called the high-dimensional GTOC (Figure 1a). The platform implants a high-dimensional vortical structure into the 2-D amplitude and phase distribution of the reflected light. The initial version of the GTOC, which includes two dielectric layers, creates a 2-D generalized parameter space determined by their thicknesses ($h_1$, $h_2$) and exhibits a non-trivial topological phase depending on the Ni layer thickness ($h_{Ni,1}$) [24]. This 2-D GTOC supports spontaneously generated OVs without structural singularities, unlike conventional devices, such as spiral phase plates [4] and metasurfaces [45, 46, 47]. The spatial distribution of the dielectric layer thicknesses bijectively maps the generalized parameter space into real space.

The high-dimensional GTOC is an advancement of the initial GTOC with additional Ni and dielectric layers, extending the dimensionality of the generalized parameter space beyond 2-D. For example, adding a pair of Ni and dielectric layers on top of the original 2-D GTOC configuration introduces the thickness of the additional dielectric layer ($h_3$) as a synthetic spatial dimension, resulting in a 3-D generalized parameter space of ($h_1$, $h_2$, $h_3$). We refer to this system as the 3-D GTOC. The two Ni layers manipulate multiple internal reflections and transmissions among the three dielectric layers, inducing complex topological interactions and topological phases. Adding further metal and dielectric layers can extend the generalized parameter space to four dimensions (4-D) and beyond.

The essential advantage of this platform is the dynamic projection of the high-dimensional generalized parameter space into 2-D real space. The additional dielectric layers consist of liquid crystal (LC) molecules with large electro-optic activity and high optical transparency. An electric bias ($V$) applied between the Ni and the optically transparent electrode (indium tin oxide, ITO) layers dynamically changes the refractive index ($n$) and the optical path length ($nh_3$) of the LC dielectric layer. The optical path length determines the underlying optical interactions across the multilayer structure. Alternatively, one can regard the thickness of the LC dielectric layer as effectively changing from $h_3$ to $h_3^* = (n_3(V)/n_3(0))h_3$ in response to the applied electric bias. Here, $n_3(0)$ and $n_3(V)$ are the refractive indices of the LC layer without and with the electric bias, respectively.

In the 3-D GTOC, the emergent vortical structure in the generalized parameter space ($h_1$, $h_2$, $h_3$) can be observed by electro-optic tomography. The electro-optic tomography projects 2-D tomographic cross-sections of generalized parameter space ($h_1$, $h_2$) into real space ($x$, $y$) while electro-optically scanning the $h_3$ axis. Consequently, the OVs in 2-D real space exhibit electrically driven dynamics depending on the bias voltage applied to the LC layer. Figure 1b shows an example of the emergent vortical structure in the generalized parameter space of the 3-D GTOC with Ni layer thicknesses ($h_{Ni,1}$, $h_{Ni,2}$) set at (20 nm, 13 nm). In the amplitude ($|r|$) and phase ($\arg(r)$) distribution of the reflected light, a pair of optical vortex and antivortex is clearly identified. Surprisingly, the optical vortex and antivortex spontaneously emerge, separate, and then fuse back together; this dynamic process originates from a vortex ring, one type of vortical structure, in the 3-D generalized parameter space. The existence, geometry, and topology of such emergent vortical structures are determined by the combination of two Ni layer thicknesses ($h_{Ni,1}$, $h_{Ni,2}$). Figure 1c shows the vortical structures, vortex rings and lines, supported by the 3-D GTOC and their transition depending on $h_{Ni,1}$. The underlying physics can be elaborated in terms of the topological phase diagram, which will be precisely discussed in Figure 2e.

Figures 1d and 1e summarize the types of emergent vortical structures of the high-dimensional GTOC depending on its dimensionality and the dynamics of the OVs projected onto 2-D real space, respectively. As dimensionality increases from 2-D to $d$-D, distinct forms of vortical structures emerge: point vortices in 2-D, vortex rings or lines in 3-D, vortex sheets in 4-D, and progressively more complex shapes in higher dimensions. The 2-D GTOC supports the spontaneous generation of real-space OVs, but fixes their positions due to the absence of synthetic spatial dimensions. The 3-D GTOC, with the addition of one electrically-driven synthetic dimension, allows OVs to move along the curved line trajectories defined by the optical vortex line or ring. In the 4-D GTOC, which supports optical vortex sheets in the generalized parametric space, electrically-driven areal dynamics of OVs are expected, depending on the external biases applied to two additional synthetic dimensions. The greater the number of active synthetic dimensions, the higher the programmability and degree of freedom of the OVs' dynamics in real space within the high-dimensional GTOC platform.

**Spontaneous generation of optical vortex line and vortex ring**

We theoretically investigate the geometry and topology of emergent vortical structures in the generalized parameter space of the 3-D GTOC (Figure 2). We employed the transfer matrix method to calculate the complex reflection coefficient ($r$) of the 3-D GTOC. For the calculations, as well as for the experimental demonstration, a $TiO_2$ layer and a glass superstrate are introduced on top of the structure in Figure 1a. The glass superstrate confines the top LC layer within a micrometer-scale gap, while the $TiO_2$ layer serves as an optical interface between the top LC layer and the glass superstrate. Details of the thickness and refractive index of each layer are described in Supplementary Notes S1 and S2. The complex reflection coefficient distribution in the 3-D generalized parameter space ($h_1$, $h_2$, $h_3$) was calculated depending on the bottom and top Ni layer thickness combinations ($h_{Ni,1}$, $h_{Ni,2}$). Here, $h_1$ and $h_2$ are the bottom and middle dielectric layer thicknesses, respectively, and $h_3$ is the top LC layer thickness. By scanning all possible combinations of ($h_{Ni,1}$, $h_{Ni,2}$), we identified two emergent vortical structures with different topologies which are the optical vortex line and optical vortex ring. On the left-hand side of Figures 2a-2d, the calculated distribution of the rotational singularities of $r = 0$ forming the emergent optical vortex line or ring in the generalized parameter space is shown. At ($h_{Ni,1}$, $h_{Ni,2}$) = (10 nm, 0 nm), (10 nm, 4 nm), and (20 nm, 9 nm), the rotational singularities are infinitely stretched as lines: the optical vortex lines (Figure 2a, 2b, 2d, left panel). At ($h_{Ni,1}$, $h_{Ni,2}$) = (10 nm, 9 nm), closed loops of the rotational singularities appear: the optical vortex rings (Figure 2c, left panel). Since the employed dielectric layers are transparent, the emergent vortical structures repeat along the $h_1$, $h_2$, and $h_3$ axes with a periodicity of $p_i = \lambda/2n_i$, where $\lambda$ is the wavelength of light and $n_i$ is the refractive index of the $i$-th dielectric layer. All calculations assume the normal incidence of 633 nm wavelength light.

The expected OV trajectories in real space are extracted from the emergent vortical structures, as shown in the right-hand side of Figure 2a-2d. The blue shaded square in the left-hand side of Figure 2a represents a 2-D slice of the generalized parameter space. When the slice is taken parallel to the ($h_1$, $h_2$)-plane, $h_3$ acts as a synthetic dimension, augmenting the two dimensions, ($x$, $y$), in real space. The intersections between the 2-D slice and the rotational singularities correspond to OVs projected in real space. Since the synthetic dimension can be electro-optically controlled, the locations of the OVs vary as a function of the voltage $V_1$ applied to the LC layer, expressed as ($x(V_1)$, $y(V_1)$). In the GTOC platform, the scale of projection between the generalized parameter space and real space is defined by the thickness gradients of the middle and bottom dielectric layers ($dx/dh_1$, $dy/dh_2$) = ($P_x/p_1$, $P_y/p_2$); $P_x$ and $P_y$ are the periodicities of the reflection coefficient in real space.

The topology and geometry of emergent vortical structures diversify the trajectories of electrically-driven real-space OV dynamics. At ($h_{Ni1}$, $h_{Ni,2}$) = (10 nm, 0 nm), diagonally distributed optical vortex lines are projected as the translational motion of real-space OVs with a winding number of $w = \pm 1$ (Figure 2a, right panel). At ($h_{Ni1}$, $h_{Ni,2}$) = (10 nm, 4 nm), the diagonally distributed helical optical vortex lines result in zig-zag type translational motion of real-space OVs (Figure 2b, right panel). At ($h_{Ni1}$, $h_{Ni,2}$) = (10 nm, 9 nm), optical vortex rings lead to the creation and annihilation of an optical vortex-antivortex pair inside the unit cell (Figure 2c, right panel). Finally, the horizontally aligned helical optical vortex lines, obtained at ($h_{Ni1}$, $h_{Ni,2}$) = (20 nm, 9 nm), cause OVs originating in a unit cell and then propagating to and annihilating in neighboring unit cells, and vice versa (Figure 2d, right panel).

The emergent vortical structures in the 3-D GTOC are classified by a topological phase diagram in the ($h_{Ni,1}$, $h_{Ni,2}$) coordinate system (Figure 2e). We inspected whether a non-trivial phase exists for each combination ($h_{Ni,1}$, $h_{Ni,2}$) and, if so, which vortical structure it supports. Non-trivial topological phases supporting emergent vortical structures are categorized into two types: vortex line

(green) and vortex ring (yellow) phases. Trivial topological phases (white) do not support any optical singularities. It should be noted that transitions between the topological phases, such as from the trivial to either of the non-trivial phases or between the two non-trivial phases themselves, involve a discontinuity. This discontinuous, sudden change in the topological properties of the system directly influences the emergent vortical structures.

The conditions supporting the non-trivial topological phases can be analyzed by the coupled resonator model with the concept of effective dimension ($d_{eff}$) (Figure 2f). Employing the Ni layers as partially reflecting/transmitting mirrors, the three dielectric layers of the 3-D GTOC act as coupled resonators. However, too thin a Ni layer cannot isolate the resonators, and too thick a Ni layer hides the existence of the resonator underneath. When $h_{Ni,1}$, the thickness of the Ni layer between the bottom and middle dielectric layers, is thinner (thicker) than a lower (upper) limit, the 3-D GTOC behaves as a 2-D GTOC of which the dimensionality is effectively reduced from three ($h_1$, $h_2$, $h_3$) to two ($h_1$, $h_2 + h_3$) or ($h_2$, $h_3$); $d_{eff} = 2$ (Fig. 2f, left panel). The singular solutions in such an effectively 2-D coordinate system are infinitely stretched along the released dimension, resulting in optical vortex lines as Figure 2a. On the other hand, in the regime where the top and bottom Ni layers function appropriately as partially reflecting/transmitting mirrors, all three resonators participate in the optical singularity generation (Fig. 2f, right panel), resulting in the vortex ring phase as Figure 2c. The creation of vortex ring phases requires three independent dimensions, a requirement that holds in any physical system. We also confirmed this understanding by the full-field simulation of the 3-D GTOC at representative points in the topological phase diagram (See Supplementary Note S3).

**Electrically-driven, versatile dynamics of real-space OVs**

We experimentally demonstrated four 3-D GTOCs with different Ni layer combinations: ($h_{Ni1}$, $h_{Ni2}$) = (10 nm, 0 nm), (10 nm, 4 nm), (10 nm, 9 nm), and (20 nm, 4 nm). These combinations support optical vortex lines and vortex rings. The samples are fabricated on $2 \times 2$ cm$^2$ Si wafers, and all layers were prepared as designed in the results in Fig. 2 (See Methods and Supplementary Note S4). The middle and bottom dielectric layers exhibit approximately ~0.25 μm thickness changes per one-millimeter distances, which maps the ($h_2$, $h_3$) space into ($x$, $y$) space in the scale of $\Delta h_i \sim 18p$ across the entire operating area ($1.5 \times 1.5$ cm$^2$) of the given sample. The top LC layer (E7 mixture, Merck), with a thickness of ~5 μm, has ordinary and extraordinary refractive indices of $n_o = 1.52$ and $n_e = 1.73$, respectively, at a wavelength of 633 nm [48]. The maximum effective thickness change induced by the applied electric bias is estimated to be $\Delta h_3 = h_3(n_e - n_o)/n_e = $ ~607 nm = ~$2.91p_3$, where $p_3 = \lambda/2n_o$. In the experiment, a sinusoidal AC electric field with a frequency of 1 kHz was applied between the Ni and ITO layers sandwiching the LC layer. The peak-to-peak voltage ($V_1$) was scanned from 0V to 20V, and the electrically driven dynamics of the real-space OVs are dramatically observed in the range of $V_1$ from 2V to 6V. The detailed operation of the LC layer, including the typical sigmoidal change of the refractive indices [41, 42], is discussed in Supplementary Note S5.

Figure 3 shows the result of electro-optic tomography, which are the amplitude and phase distributions of the complex reflection coefficient ($r$) depending on the electric bias applied to the LC layer (See Methods). The locations of the OVs are indicated by the red and blue circles in the bottom panel of Figures 3a-3d. For the sample of ($h_{Ni1}$ = 10 nm, $h_{Ni2}$ = 0 nm), the optical vortex-antivortex pair exhibits translational motion along the trajectories of straight lines (Figure 3a). In the sample of ($h_{Ni1}$ = 10 nm, $h_{Ni2}$ = 4 nm), the optical vortex and antivortex move along zigzag-shaped trajectories (Figure 3b). The sample of ($h_{Ni1}$ = 10 nm, $h_{Ni2}$ = 9 nm) demonstrates the creation and annihilation of the optical vortex-antivortex pair at $V_1$ of 2.68V and 3.04V, respectively, revealing the optical vortex ring (Figure 3c). The sample of ($h_{Ni1}$ = 20 nm, $h_{Ni2}$ = 9 nm) causes more complex dynamics of the OVs originating from the helical optical vortex line (Figure 3d). The optical vortex-antivortex pairs are created at 3.00V and annihilated at 3.12V and 3.16V. In the annihilation process, the optical vortex and antivortex come from adjacent unit cells merge together. The measured trajectories of the real-space OVs well agree with the theoretical estimations in Figures 2a-2d.

**Topological transitions of emergent 3-D vortical structures**

We confirm that the observed real-space OV dynamics originate from the optical vortex line and optical vortex ring structures, which can be directly visualized in the ($x$, $y$, $V_1$) space (Figure 4). We prepared ten samples (Samples A-J) with different combinations of ($h_{Ni,1}$, $h_{Ni,2}$) along two distinct paths in the topological phase diagram (the black dashed lines in Figure 2e). Samples A, B, C, D, and E have different top Ni layer thicknesses as $h_{Ni,2}$ = 0, 4, 6, 9, and 15 nm, respectively, with the same bottom Ni layer thickness of $h_{Ni,1}$ = 10 nm. Samples F, G, H, I, and J share the same sequence of top Ni layer thicknesses but with $h_{Ni,1}$ = 20 nm. The electrically driven dynamics of the real-space OVs depending on the applied electric bias ($V_1$) on the top LC layer clearly reconstruct the vortical structures in the 3-D generalized parameter space. Here, samples A, B, D, and I correspond to the theoretical cases in Figures 3a-3d. As shown in Fig. 4a, the real-space OV dynamics from samples A to E correlate to the vortex line (A), helical vortex line (B), and vortex ring (C, D, and E). In contrast, from samples F to J, the real-space OV dynamics originate from the trivial phase without any singularities (F), vortex ring (G), helical vortex line (H, I), and vortex ring (J), respectively (Fig. 4c). The measured distributions of $|r|$ and arg($r$) in real space from samples C, E, F, G, H, I, and J are detailed in Supplementary Note S6.

The measured vortical structures show good agreement with the theoretical estimations. Figure 4b and 4d are the calculated topological phase diagrams for $h_{Ni,1}$ = 10 nm and 20 nm, respectively, which are depicted as the black dashed lines in Figure 2e. The locations of experimental sample conditions are indicated by light gray dashed lines in the phase diagram. From sample A to J, the geometry and topology of measured vortical structures are consistent with the topological phase diagrams as well as the calculated singularity distributions in the generalized parameter space, which are shown in Supplementary Note S7. Note that the emergent 3-D vortical structures can be a building block in the construction of complex vortical structures. An introduction of additional optical anisotropy in 3-D GTOC can lead to the generation of vortex braiding and vortex links, which are discussed in Supplementary Note S8.

**Proposal of emergent 4-D vortical structures**

We theoretically investigate and propose emergent vortical structures beyond the 3-D generalized parameter space. Here, we consider a 4-D GTOC with a generalized parameter space consisting of two dielectric layers with gradient thicknesses ($h_1$, $h_2$) and two LC layers with thicknesses ($h_3$, $h_4$). The topological phase diagram extends to 3-D as a function of the thicknesses of three Ni layers ($h_{Ni,1}$, $h_{Ni,2}$, $h_{Ni,3}$) and determines the geometry and topology of emergent 4-D vortical structures. Figure 5a shows representative calculated emergent vortical structures in the 4-D GTOC visualized by 3-D coordinates ($h_1$, $h_2$, $h_3$) and 1-D color coordinate ($h_4$). Three different combinations of the Ni layers' thicknesses ($h_{Ni,1}$, $h_{Ni,2}$, $h_{Ni,3}$) give rise to planar vortex sheets (10 nm, 0 nm, 2 nm), hollow vortex cylinders (10 nm, 10 nm, 0 nm), and complex vortex manifold (10 nm, 10 nm, 2 nm), respectively.

The OV trajectories from the emergent optical vortex sheets are expected to show versatile areal dynamics in real space (Figures 5b and 5c). Here, similar to the results in Figures 2a-2d, the locations of optical singularities are extracted from 2-D slices in the generalized parameter space. The 2-D slices, parallel to the ($h_1$, $h_2$) plane, utilize $h_3$ and $h_4$ as two different synthetic dimensions. If $h_3$ and $h_4$, representing the thicknesses of two LC layers, can be effectively changed by the electric biases of $V_1$ and $V_2$, the trajectories of the real-space OVs in the 4-D GTOC can be controlled with increased degrees of freedom. Figure 5b (and 5c) shows the dynamic trajectories of the optical vortex (red areas and lines) and the antivortex (blue areas and lines), for the scenarios where $h_3$ varies and $h_4$ is fixed (and vice versa). For the planar vortex sheets ($h_{Ni,1}$, $h_{Ni,2}$, $h_{Ni,3}$) = (10 nm, 0 nm, 2 nm), translational motions of real-space OVs are expected under $V_1$ and $V_2$. For the cylindrical optical vortex sheet (10 nm, 0 nm, 10 nm), the $V_1$ and $V_2$ induce distinct real-space OV dynamics; (1) translational motions under $V_1$ and (2) creation/annihilation of optical vortex-antivortex pair in a closed loop under $V_2$. For complex folded vortex sheet (10 nm, 10 nm, 2 nm), the creation/annihilation of optical vortex-antivortex pair in closed loops are expected under both $V_1$ and $V_2$. Note that the trajectories of OVs cover extensive areas of the unit cell (red and blue shaded regions), which implies a possibility of programmable manipulation of real-space OV dynamics by $V_1$ and $V_2$.

**Discussion**

We theoretically and experimentally investigate emergent optical vortical structures in high dimensions and their dynamic projections, specifically OVs, onto 2-D real space. The high-dimensional GTOC incorporates multiple synthetic dimensions, corresponding to its number of dielectric layers. We realize various high-dimensional optical vortical structures, such as optical vortex lines, rings, and sheets, and explore their topological transitions, which depend on the coupling properties among the multilayers of the GTOC. Based on electro-optic tomography, we demonstrate electrically-driven, dynamic real-space OVs, whose trajectories faithfully reconstruct the geometry and topology of the intrinsic vortical structures. We believe our work opens up a new direction for emulating and exploring high-dimensional vortical structures and their topological interactions through convenient optical methods. We anticipate that this exploration could extend to high-dimensional Weyl physics [48, 49], topological pumping [50], non-abelian Yang monopole [51], and quantum Hall effect [52], which have mainly been investigated by theoretical approaches. Employing optically dispersive media could enable the investigation of toroidal vortices [53, 54], spatiotemporal optical vortices [55, 56], and optical skyrmions [57]. Furthermore, simultaneous manipulation on the multiple degrees of freedom of light, such as polarization, wavelength, and spatiotemporal waveform, will be further developed based on anisotropic and nonlinear optical responses [58,59,60,61,62].

## Methods

### Fabrication of 3-D GTOC

We prepared a 2 cm × 2 cm Si substrate on which a 100 nm thick Al mirror layer was deposited by electron beam evaporation. The height-varying bottom and middle dielectric layers were made of optical adhesive polymer (NOA 6311), which was injected into a wedge-shaped sandwich cell at 90°C and irradiated with a 365-nm wavelength UV lamp for 30 minutes (Supplementary Note S4). After UV curing, the top cladding of the sandwich cell was removed for the next step. Between the fabrication of the bottom and middle dielectric layers, the sample was rotated by 90° so that the thickness gradients of the two layers are nearly perpendicular. The top liquid crystal layer consists of E7 mixture (Merck). For the electro-optic modulation, a sandwich cell of a 1 mm thick glass superstrate was placed on the top Ni layer. Then, a 100 nm thick indium tin oxide (ITO) layer and a 15 nm thick $TiO_2$ layer were deposited on the glass superstrate by electron beam evaporation. A planar anchoring polyimide (PAPI, Dow Chemical Co.) layer is prepared on top of the $TiO_2$ layer, accompanied by a mechanical rubbing process to align LC molecules. The bottom and top Ni layers with uniform thicknesses were deposited by electron beam evaporation.

### Measurement Set-Up

The off-axis holography set-up is implemented as described in Ref. [24]. To utilize the linear birefringence of LC, the incident light is linearly polarized and aligned parallel to the direction of mechanical rubbing. The electrical activation of molecule reorientation in the LC layer is facilitated using an arbitrary waveform generator (DG4162, RIGOL) and a broadband linear amplifier (A400, Pendulum). We apply a 1-kHz sinusoidal wave to the LC layer through two copper wires soldered to the surfaces of the Ni layer and the ITO layer, respectively. When the top Ni layer is absent ($h_{Ni,2} = 0$ nm), the electric bias is applied between the bottom Ni and ITO layer, which requires approximately 2 times larger voltages for the molecule alignment compared to the $h_{Ni1} \neq 0$ cases. Prior to measurement, the molecule alignment of the LC layer along the direction of mechanical rubbing in the PAPI layer is initialized by thermal annealing at 90°C for 1 minute.

### Data Availability

The data that support the findings of this study are available from the corresponding authors upon reasonable request.

**Code Availability**

All codes generated during this study are available upon request from the corresponding authors.

**Acknowledgement**

D.K. acknowledges the support of the National Research Foundation of Korea (NRF, RS-2023-00240304). J.S. acknowledges the support of the NRF (RS-2024-00414119). S.F. acknowledges the support of a MURI project from the U. S. Air Force Office of Scientific Research (Grant No. FA9550-22-1-0339), and the U. S. Army Research Office (Grant No. W911NF-24-2-0170). D.Y. acknowledges the support of the NRF (2021M3H4A3A01050378). M.-K.S. acknowledges the support of the National Research Foundation of Korea (RS-2024-00350185 and RS-2024-00408271). This work was performed, in part, at the Center for Integrated Nanotechnologies, an Office of Science User Facility operated for the U.S. Department of Energy (DOE) Office of Science. Los Alamos National Laboratory, an affirmative action equal opportunity employer, is managed by Triad National Security, LLC for the U.S. Department of Energy's NNSA, under contract 89233218CNA000001.

**Author Contribution**

D. K. conceived the idea. G. P., D. K., Y. C., A. B., J. H., and H. Y. fabricated the 3-D GTOC samples. D. K. performed the theoretical calculations of 3-D and 4-D GTOCs and the optical characterization of the fabricated 3-D GTOC samples. D. K. and M.-K.S. analysed the data and wrote the manuscript with input from all other authors.

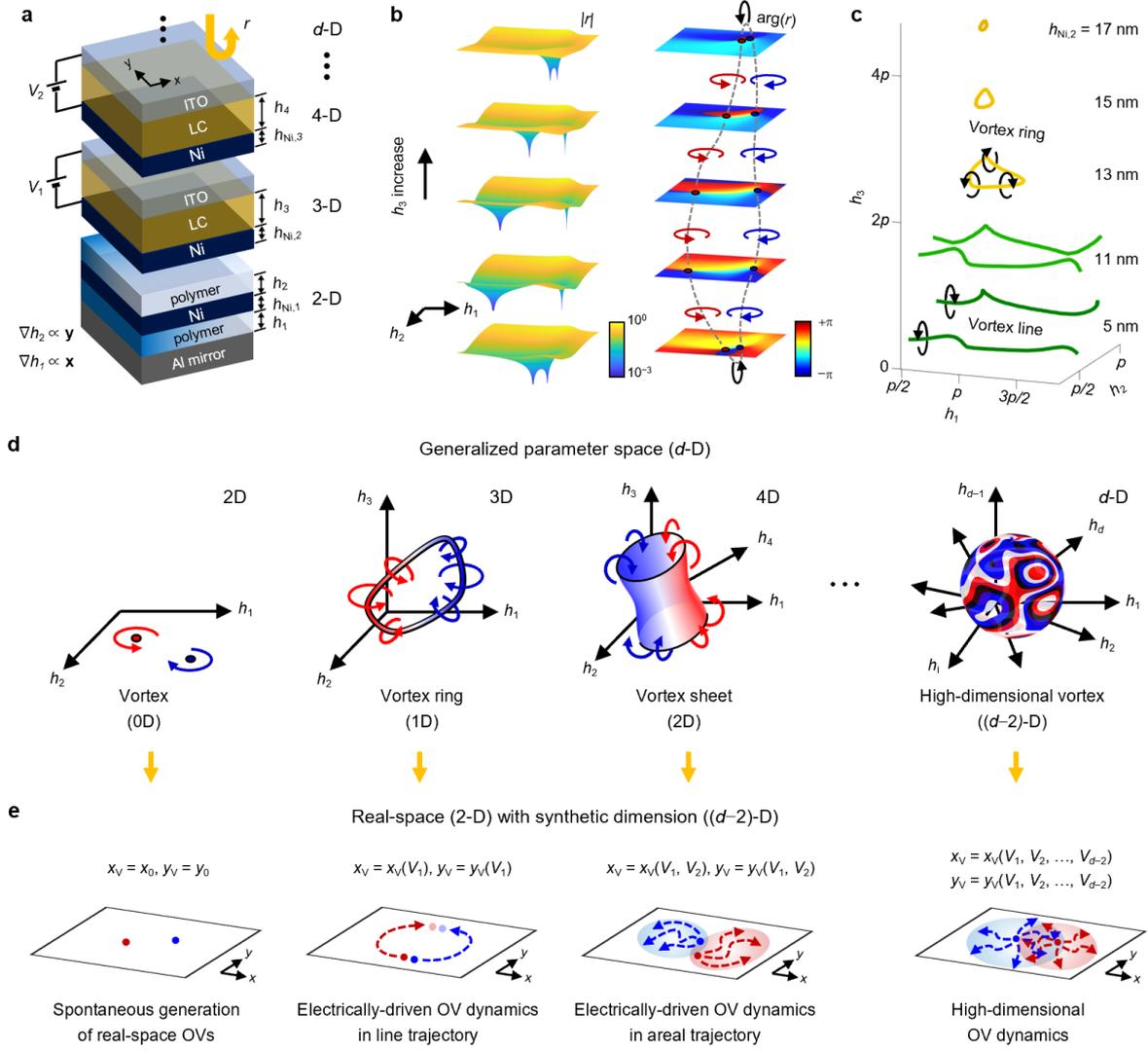

**Figure 1 Emergent vortical structure generation in high-dimensional GTOC. (a)** Schematic of high-dimensional GTOC consisting of polymer, liquid crystal (LC), ITO, and Ni layers. The thicknesses of two polymer layers ($h_1$, $h_2$) vary in $x$ and $y$ direction, respectively. Electric biases ($V_1, V_2, \ldots$) applied between the ITO and Ni layers induce effective changes of the LC layer thicknesses ($h_3, h_4, \ldots$). The thicknesses of the Ni layers ($h_{Ni,1}$, $h_{Ni,2}$, $h_{Ni,3}$, …) determine the emergence and topology of optical vortices. The dimensionality ($d = 2, 3, \ldots$) of the high-dimensional GTOC is defined by the number of dielectric (polymer and LC) layers. **(b)** Amplitude ($|r|$) and phase ($\arg(r)$) distributions of the reflected light from a 3-D GTOC ($h_{Ni,1} = 20$ nm, $h_{Ni,2} = 13$ nm), supporting an optical vortex ring in the generalized parameter space ($h_1, h_2, h_3$). **(c)** Topological transition from an optical vortex line to ring as $h_{Ni,2}$ changes from 5 nm to 17 nm. The periodicity ($p_i$) is given as $\lambda/2n_i$, where $n_i$ is the refractive index of the $i$-th dielectric layer. **(d,e)** Emergent vortical structures in the generalized parameter space supported by the high-dimensional GTOC and the dynamics of OVs' motions in real space, depending on its dimensionality.

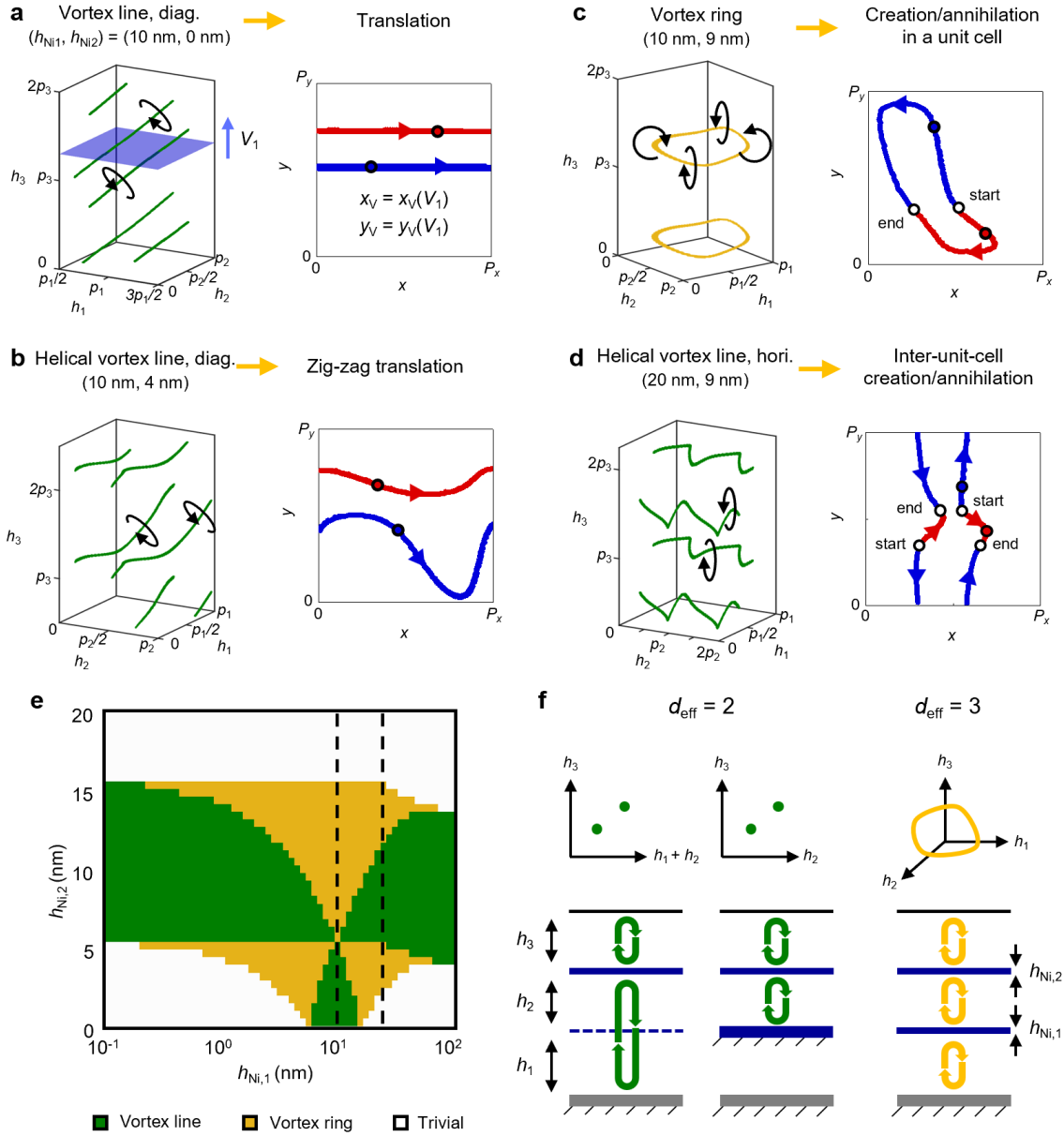

**Figure 2 Emergent optical vortex lines and vortex rings in 3-D GTOC.** **(a-d)** Calculated distributions of the rotational singularities ($r = 0$) in the generalized parameter space (left panel) and their dynamic motions in real space (right panel). **(a)** The 3-D GTOC with ($h_{Ni,1}$, $h_{Ni,2}$) = (10 nm, 0 nm) supports optical vortex lines and causes the OVs to move in a translational motion. The blue shaded square in the left panel indicates a slice of the generalized parameter space at a fixed $h_3$ that can be electro-optically scanned by $V_1$ (blue arrow). **(b)** The 3-D GTOC with ($h_{Ni,1}$, $h_{Ni,2}$) = (10 nm, 4 nm) induces helical optical vortex lines and makes the OVs move in a zig-zag motion. **(c)** The emergent optical vortex rings by the 3-D GTOC with ($h_{Ni,1}$, $h_{Ni,2}$) = (10 nm, 9 nm) cause the generation and annihilation of an optical vortex-antivortex pair along a closed-loop trajectory. **(d)** The 3-D GTOC with ($h_{Ni,1}$, $h_{Ni,2}$) = (20 nm, 9 nm) supports helical optical vortex lines and causes the generation and annihilation of optical vortex-antivortex pairs across adjacent unit cells. **(e)** Topological phase diagram

of the 3-D GTOC depending on $h_{Ni,1}$ and $h_{Ni,2}$. The non-trivial topological phases of optical vortex lines and rings are supported in the green and yellow regions, respectively. Trivial phases without singularities appear in the white regions. The dashed black lines indicate the conditions of the experimental demonstration. **(f)** Coupled resonator model of the 3-D GTOC. The model highlights the effect of the bottom Ni layer's thickness: when too thick or thin, it reduces the system's effective dimension ($d_{eff}$) to two; a moderate thickness maintains an effective dimensionality of three.

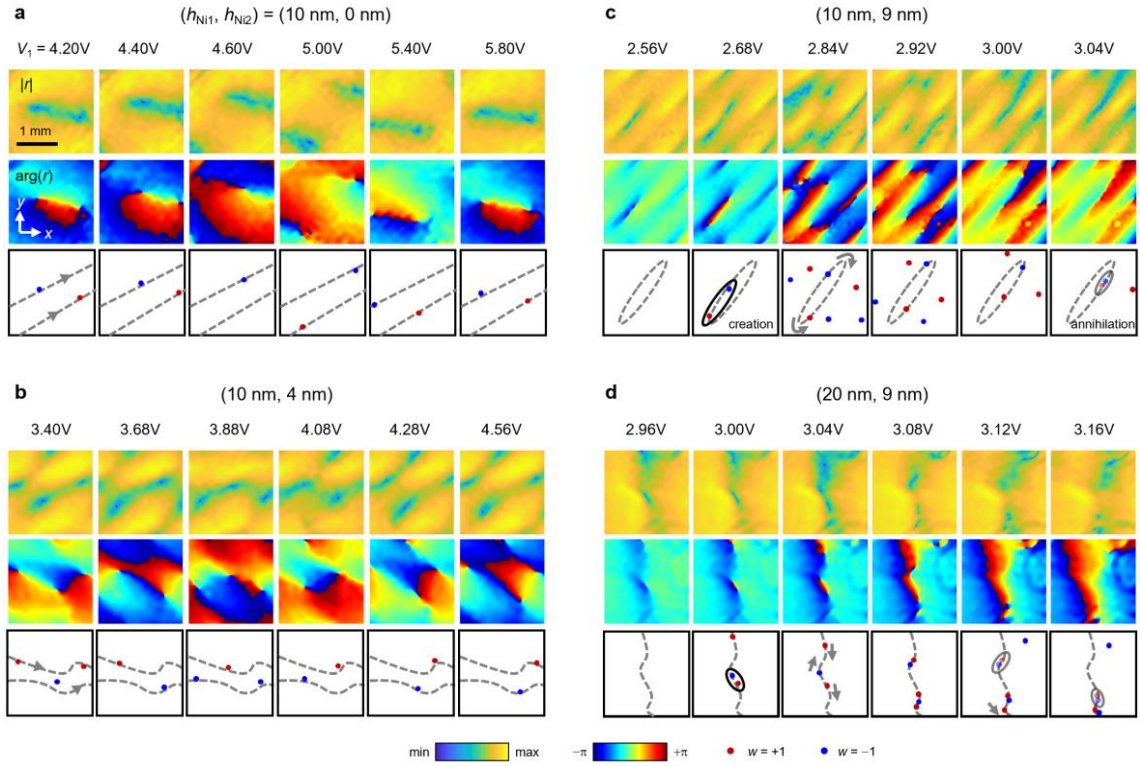

**Figure 3 Electrically-driven dynamics of real-space OVs. (a-d)** Measured amplitude ($|r|$) and phase ($\arg(r)$) distributions of the complex reflection coefficient of four representative 3-D GTOCs depending on the applied electric bias ($V_1$). The positions of the real-space OVs are extracted from the measured phase distribution (bottom panel). The red and blue dots indicate the optical vortex ($w = +1$) and antivortex ($w = -1$), respectively. The gray dashed lines guide the trajectories of the OVs in real space. **(a)** Translational motion of OVs for the configuration $(h_{Ni,1}, h_{Ni,2}) = (10\text{ nm}, 0\text{ nm})$. **(b)** Zig-zag type translational motion of OVs for $(h_{Ni,1}, h_{Ni,2}) = (10\text{ nm}, 4\text{ nm})$. **(c)** Generation and annihilation of an optical vortex-antivortex pair along a closed loop for $(h_{Ni,1}, h_{Ni,2}) = (10\text{ nm}, 9\text{ nm})$. **(d)** Generation and annihilation of optical vortex-antivortex pairs across adjacent unit cells for $(h_{Ni,1}, h_{Ni,2}) = (20\text{ nm}, 9\text{ nm})$. Scale bar is 1 mm.

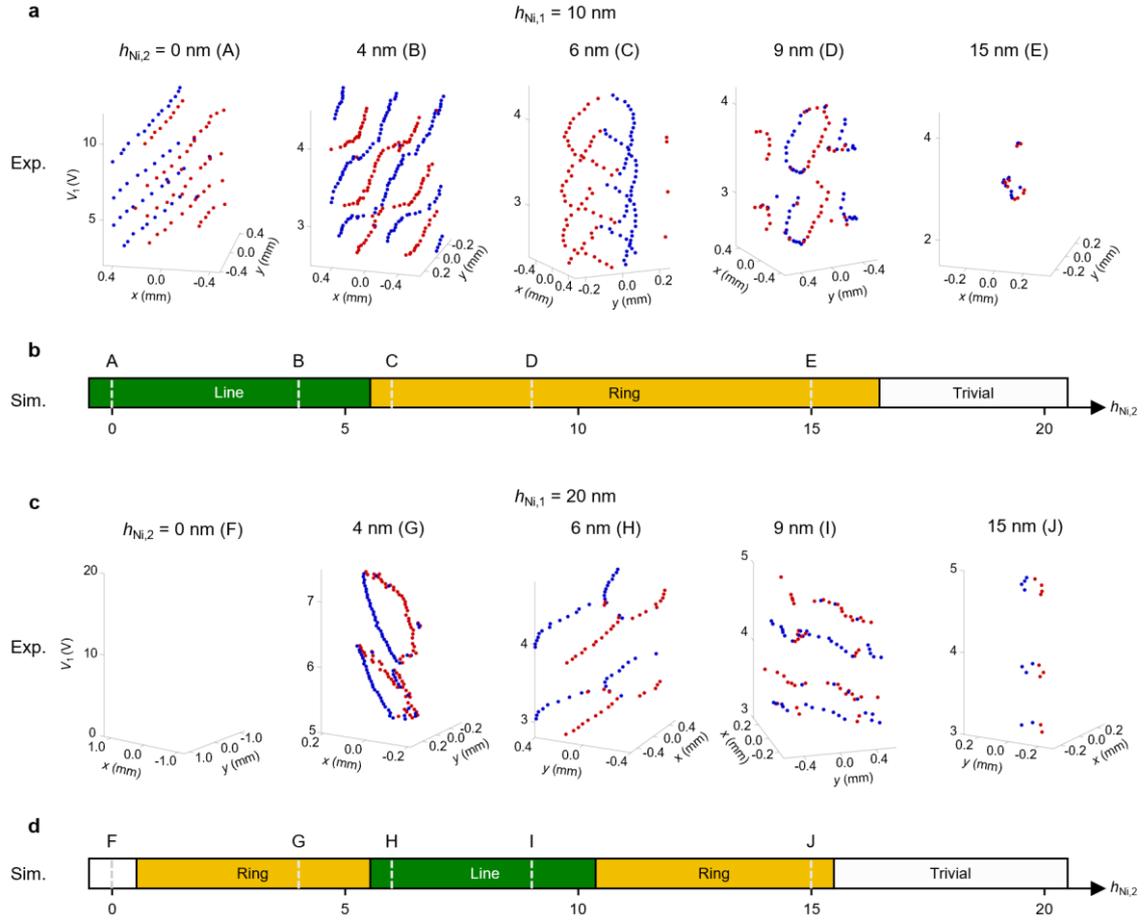

**Figure 4 Topological phase transitions in 3-D GTOCs. (a,c)** Measured trajectories of electrically-driven real-space OV dynamics in ten different 3-D GTOCs: **(a)** Samples A-E with $h_{Ni,1}$ = 10 nm and **(c)** samples F-J with $h_{Ni,1}$ = 20 nm. The plots in $(x, y, V_1)$ coordinates visualize the distributions of the rotational singularities in the generalized parameter space, which agree well with the theoretical calculations shown in Fig. 2. The red and blue circles correspond to the optical vortex ($w$ = +1) and antivortex ($w$ = −1), respectively. **(b, d)** Calculated topological phase diagram as a function of $h_{Ni,2}$ for **(b)** $h_{Ni,1}$ = 10 nm and **(d)** $h_{Ni,1}$ = 20 nm (the black dashed lines in Fig. 2e). The light gray dashed lines indicate the positions of samples A-J.

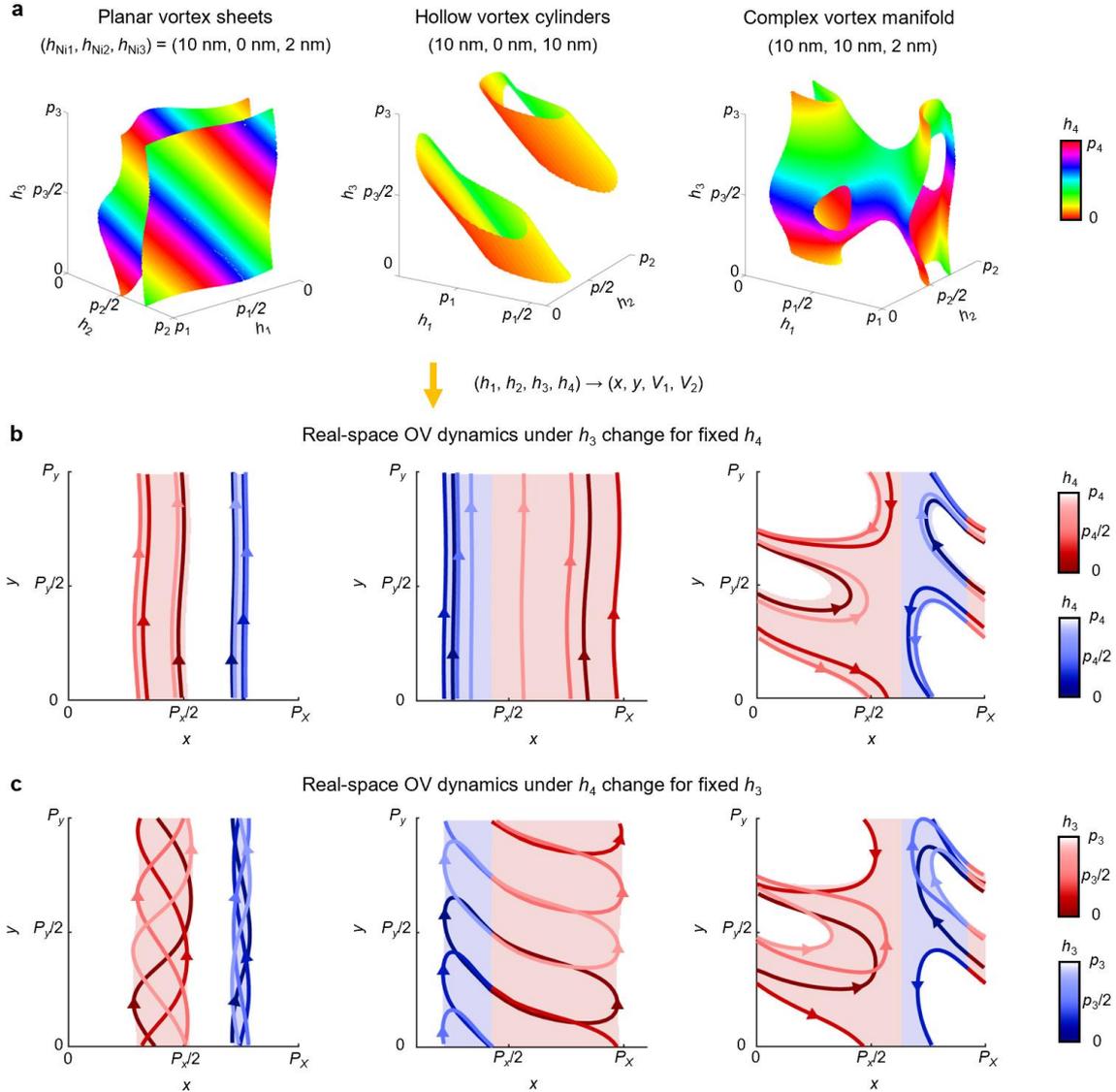

**Figure 5 Emergent optical vortex sheets in 4-D GTOCs. (a)** Calculated rotational singularity distributions of emergent optical vortex sheets for three different 4-D GTOCs. Displayed from left to right are: a pair of optical vortex sheets at $(h_{Ni,1}, h_{Ni,2}, h_{Ni,3})$ = (10 nm, 0 nm, 2 nm), a cylindrical vortex sheet at (10 nm, 0 nm, 10 nm), and a complex folded vortex sheet at (10 nm, 10 nm, 2 nm). To visualize the vortical structures in the 4-D generalized parameter space, the thickness of the fourth dielectric layer ($h_4$) is represented by a hue color scale. **(b,c)** Trajectories of real-space OV dynamics originating from the optical vortex sheets with **(b)** varying $h_4$ for a fixed $h_3 = 0$, $p_3/4$, $p_3/2$, and $3p_3/4$, and **(c)** varying $h_3$ for a fixed $h_4 = 0$, $p_4/4$, $p_4/2$, and $3p_4/4$. The reddish and bluish solid lines represent the trajectories of the optical vortex and antivortex, respectively, with different values of the variable ($h_4$ or $h_3$) corresponding to the color bar. The red and blue shaded regions indicate the areas where the optical vortex and antivortex are projected, respectively.